\documentclass[aps,prl,twocolumn,superscriptaddress]{revtex4-1}
\usepackage{graphicx}
\usepackage{color}
\usepackage{amsfonts}
\newcommand{\rev}{\textcolor{black}}

\begin{document}

\title{Shear-induced anisotropy in rough elastomer contact}

\author{R. Sahli}
\affiliation{Univ Lyon, Ecole Centrale de Lyon, ENISE, ENTPE, CNRS, Laboratoire de Tribologie et Dynamique des Syst\`emes LTDS, UMR 5513, F-69134, Ecully, France}
\author{G. Pallares}
\affiliation{Univ Lyon, Ecole Centrale de Lyon, ENISE, ENTPE, CNRS, Laboratoire de Tribologie et Dynamique des Syst\`emes LTDS, UMR 5513, F-69134, Ecully, France}
\affiliation{CESI, LINEACT, Zone A\'eroportuaire M\'editerran\'ee, 34130 Mauguio, France}
\author{A. Papangelo}
\affiliation{Dipartimento di Meccanica, Matematica e Management, Politecnico di Bari, Viale Japigia 182, 70126 Bari, Italy}
\affiliation{Hamburg University of Technology, Department of Mechanical Engineering, Am Schwarzenberg-Campus 1, 21073 Hamburg, Germany}
\author{M. Ciavarella}
\affiliation{Dipartimento di Meccanica, Matematica e Management, Politecnico di Bari, Viale Japigia 182, 70126 Bari, Italy}
\affiliation{Hamburg University of Technology, Department of Mechanical Engineering, Am Schwarzenberg-Campus 1, 21073 Hamburg, Germany}
\author{C. Ducottet}
\affiliation{Universit\'e de Lyon, UJM-Saint-Etienne, CNRS, IOGS, Laboratoire Hubert Curien UMR5516, F-42023, Saint-Etienne, France}
\author{N. Ponthus}
\affiliation{Univ Lyon, Ecole Centrale de Lyon, ENISE, ENTPE, CNRS, Laboratoire de Tribologie et Dynamique des Syst\`emes LTDS, UMR 5513, F-69134, Ecully, France}
\author{J. Scheibert}
\affiliation{Univ Lyon, Ecole Centrale de Lyon, ENISE, ENTPE, CNRS, Laboratoire de Tribologie et Dynamique des Syst\`emes LTDS, UMR 5513, F-69134, Ecully, France}
\email{julien.scheibert@ec-lyon.fr}

\date{\today}

\begin{abstract}
True contact between randomly rough solids consists of myriad individual micro-junctions. While their total area controls the adhesive friction force of the interface, other macroscopic features, including viscoelastic friction, wear, stiffness and electric resistance, also strongly depend on the size and shape of individual micro-junctions. Here we show that, in rough elastomer contacts, the shape of micro-junctions significantly varies as a function of the shear force applied to the interface. This process leads to a growth of anisotropy of the overall contact interface, which saturates in macroscopic sliding regime. We show that smooth sphere/plane contacts have the same shear-induced anisotropic behaviour as individual micro-junctions, with a common scaling law over four orders of magnitude in initial area. \rev{We discuss the physical origin of the observations in the light of a fracture-based adhesive contact mechanics model, described in the companion article, which captures the smooth sphere/plane measurements}. Our results shed light on a generic, overlooked source of anisotropy in rough elastic contacts, not taken into account in current \rev{rough} contact mechanics models.

\end{abstract}


\maketitle

Real contact between rough solids only occurs in randomly distributed small regions of the interface (micro-junctions)~\cite{greenwood_contact_1966,persson_theory_2001}.
The adhesion component of the friction force is proportional to the total area of all micro-junctions~\cite{archard_elastic_1957, dieterich_direct_1994,wu-bavouzet_effect_2010,degrandi-contraires_sliding_2012, sahli_evolution_2018, weber_molecular_2018}.
In contrast, many other macroscopic contact properties (\textit{e.g.} electric and heat resistance~\cite{greenwood_constriction_1966,popov_contact_2017}, normal and shear stiffnesses~\cite{medina_analytical_2013}, wear~\cite{aghababaei_debris-level_2017} and viscoelastic friction~\cite{scaraggi_friction_2015}) also depend on the details of the real contact morphology, including the number, size, spatial distribution and shape of individual micro-junctions. In this context, it is clear that any phenomenon affecting the real contact morphology of an interface will also affect all of its above-mentioned physical properties.

The real contact morphology of rough interfaces depends both on loading (\textit{e.g.} pressure and contact time~\cite{dieterich_imaging_1996} and sliding velocity~\cite{carbone_rough_2014}) and system parameters (\textit{e.g.} adhesion between the solids~\cite{pastewka_contact_2014} and the spectral contents of the surface roughness~\cite{bush_elastic_1975,scaraggi_friction_2015,yastrebov_role_2017})~\cite{vakis_modeling_2018}. One of the important descriptors of morphology is the degree of anisotropy of the interface. Rough contact anisotropy may occur for various reasons: anisotropic roughness~\cite{carbone_contact_2009}, anisotropic bulk material behaviour or viscoelasticity in gross sliding regime~\cite{carbone_rough_2014}. The very same reasons also yield anisotropic contact shapes at the interface between smooth axisymmetric bodies, for instance in sphere/plane geometry~\cite{johnson_contact_1987,barber_JKR_2014,koumi_rolling_2015, fretigny_contact_2017}, thus suggesting common physical origins.

Interestingly, in such smooth sphere/plane contacts, another source of anisotropy has been observed: an initially circular contact becomes less and less axisymmetric as it is increasingly sheared~\cite{savkoor_effect_1977,varenberg_shearing_2007,petitet_materiaux_2008,waters_mode-mixity-dependent_2010,sahli_evolution_2018, mergel_continuum_2019}. It is thus natural to hypothesize that a similar growth of anisotropy may also occur in multi-contact interfaces under shear. Such a behaviour would imply that many transport, mechanical or tribological properties of a rough contact are not intrinsic features of the interface but are actually dynamical quantities that evolve with the amount of shear applied. In order to test this hypothesis, we further analyze an extensive series of experiments performed on various elastomeric multi-contacts, in which the evolution of the morphology of the real contact is monitored optically as the shear force is increased from pure normal contact to gross sliding. All experimental details can be found in~\cite{sahli_evolution_2018}, while the main points are summarized here.

\begin{figure}[ht]
\includegraphics[width=\columnwidth]{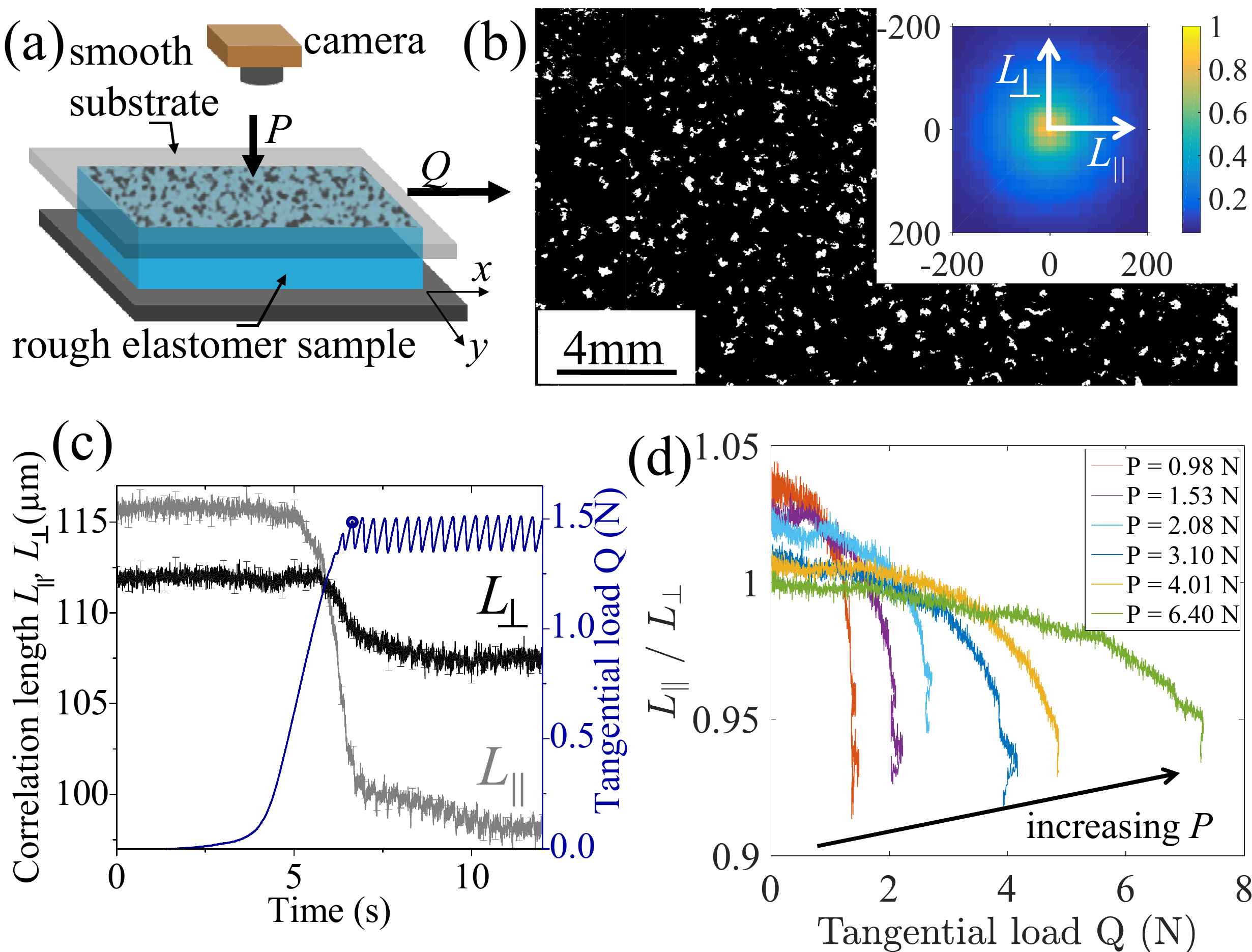}
\caption{(a) Sketch of the experimental setup. (b) Typical binarized image of a multicontact. $P$=6.40N. Inset: normalized autocorrelation function of the image shown in main panel. The correlation lengths $L_{\parallel}$ and $L_{\perp}$ are obtained through fitting with $e^{-\sqrt{x^2/{L_{\parallel}}^2+y^2/{L_{\perp}}^2}}$ \cite{Sahli_SMPRL_2019}. (c) Concurrent time-evolutions of $Q$, $L_{\parallel}$ and $L_{\perp}$ for the interface of (b). $\circ$: static friction peak $Q_s$. Error bar: 95\% confidence interval. (d) Evolution of $L_{\parallel}/L_{\perp}$ \textit{vs} $Q$, for various normal forces.\label{Fig1}}
\end{figure}

We make centimeter-sized contact (Fig.~\ref{Fig1}a) between a flat smooth bare glass slider and a flat rough crosslinked polydimethylsiloxane (PDMS) block (\textit{rms} roughness 26$\mu$m, \rev{see a typical Power Spectrum Density in~\cite{Sahli_SMPRL_2019}}), under constant normal force $P$ in the range \rev{0.98-6.40~N (\textit{i.e.} a ratio $p/E^*$ in the range 0.0006-0.0043, with $p$ the average pressure applied, $E^*=E/(1-\nu^2)$, and $E$ and $\nu$ the Young's modulus and Poisson's ratio of the PDMS, respectively)}.  We optionally coat the glass surface, either with grafted PDMS chains or with a layer of cross-linked PDMS, to change its adhesive and frictional properties. We drive the slider horizontally towards macrosopic sliding, at constant velocity $V$ in the range 0.05-1~mm/s, while we monitor the tangential force $Q$. Simultaneously, we image \textit{in situ} the contact and access highly contrasted pictures, which we efficiently binarize using automatic thresholding~\cite{Sahli_SMPRL_2019}, enabling identification of each micro-junction (white spots in Fig.~\ref{Fig1}b). In the following, we only present data for PDMS/glass interfaces: PDMS/crosslinked-PDMS interfaces behave similarly, despite a slight initial anisotropy, while PDMS/grafted-PDMS interfaces show negligible evolution under shear due to low frictional strength~\cite{sahli_evolution_2018}. Also, we observed only a weak effect of $V$, so we will only show results for $V$=0.1mm/s.

To assess whether the degree of anisotropy of our rough contacts changes under shear, we compute, for each binarized image, its normalized autocorrelation function, a typical example of which is shown on Fig.~\ref{Fig1}b (inset). We then fit this function (caption of Fig.~\ref{Fig1}) to extract two correlation lengths, $L_{\parallel}$ and $L_{\perp}$, \rev{in the directions parallel and orthogonal to shear loading}, respectively. Figure~\ref{Fig1}c shows a typical concurrent evolution of the correlation lengths and the tangential force $Q$, as the interface is driven from its initial state (pure normal force) to macroscopic sliding. We find that, for $Q=0$, $L_{\parallel}$ and $L_{\perp}$ are roughly equal (less than 5$\%$ difference), for all normal forces, showing that the contact is initially isotropic. As soon as $Q$ increases, $L_{\parallel}$ is found to significantly decrease, by typically 10-15$\%$, while $L_{\perp}$ varies much less. Both correlation lengths stabilize after $Q$ has reached the static friction peak value, $Q_s$, and the interface has entered a macroscopic sliding regime. We quantify contact anisotropy by the ratio $L_{\parallel}/L_{\perp}$, shown as a function of the tangential force $Q$ in Fig.~\ref{Fig1}d. We find that $L_{\parallel}/L_{\perp}$ decreases by $\sim$7-12$\%$, with larger decays for smaller normal forces. Those results validate our initial hypothesis (rough contacts undergo a growing anisotropy under increasing shear) with, at the onset of sliding, a significantly reduced characteristic length scale of the real contact along the loading direction.

\begin{figure}[ht]
\includegraphics[width=\columnwidth]{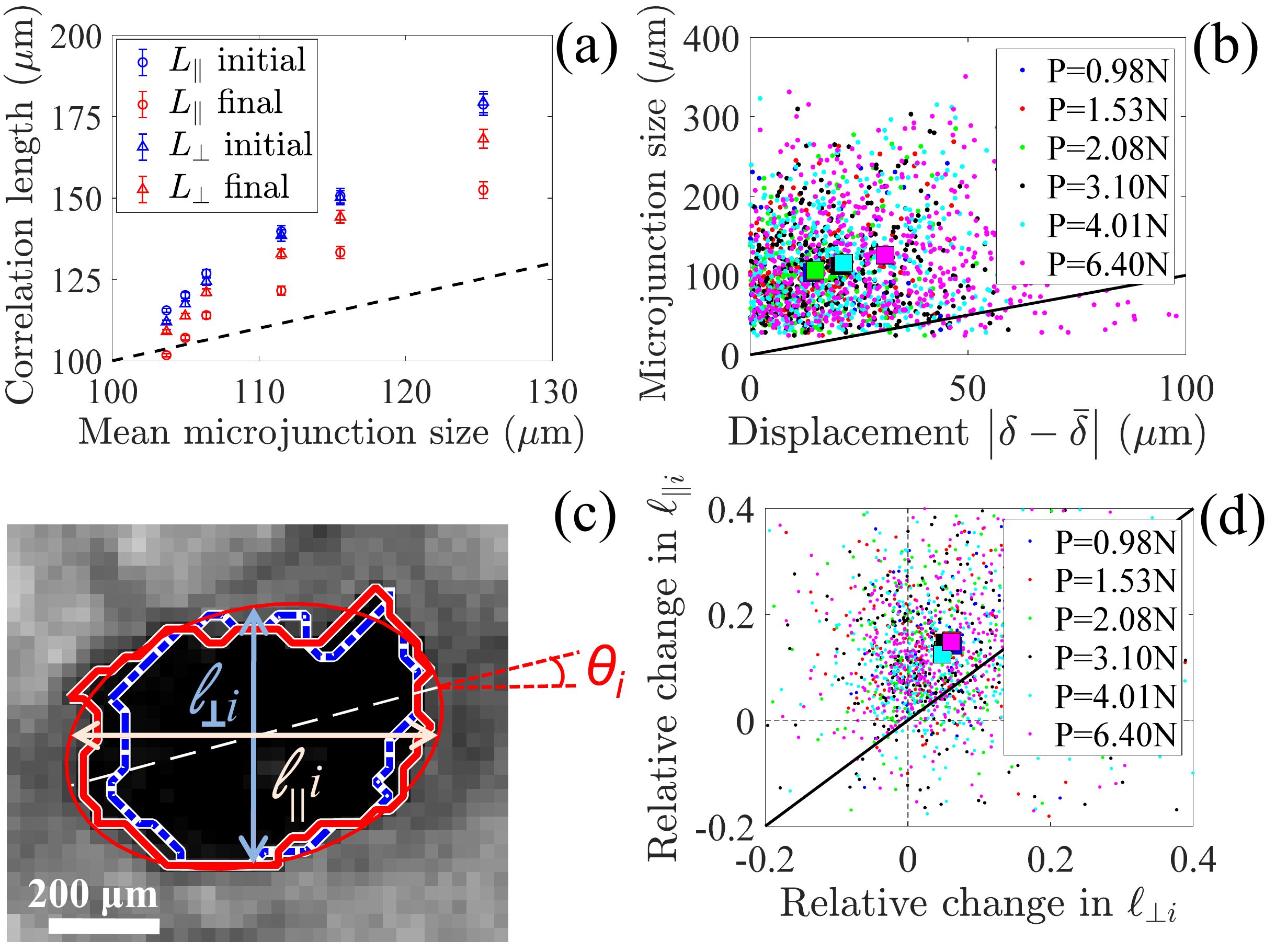}
\caption{(a) Macroscopic correlation lengths along \rev{the directions parallel ($\circ$) and orthogonal ($\triangle$) to shear}, for $Q=0$ (blue) and $Q=Q_s$ (red), \textit{vs} average micro-junction size, for various $P$. Error bar: 95\% confidence interval. (b) $\bullet$: Micro-junction size \textit{vs} norm of its excursion from the initial real contact pattern, for the experiments of (a). $\square$: data barycenter for each $P$. Solid line: equality line. (c) Typical micro-junction image. Solid red (dashed blue) line: contour for $Q$=0 ($Q$=$Q_s$). Red ellipse: \rev{equivalent} ellipse, showing $\ell_{\parallel i}$, $\ell_{\perp i}$ and $\theta_i$. (d) Relative change of $\ell_{\parallel i}$ \textit{vs} relative change of $\ell_{\perp i}$, for individual micro-junctions in the experiments of (a) and (b). $\square$: data barycenter, as in (b). Solid line: equality line.\label{Fig2}}
\end{figure}

What is the microscopic origin of this growing anisotropy? To answer this question, we track the individual micro-junctions along the shearing experiments (tracking performed as in~\cite{sahli_evolution_2018}) and extract the time-evolution of their area and shape and the location of their center of mass. For each tracked micro-junction $i$, we define its \rev{mean} size $a_i$ from its area $A_i$: $a_i=\sqrt{A_i/\pi}$. We first find that, for all experiments, the values of $L_{\parallel} $ and $L_{\perp}$ match the \rev{mean} size of individual micro-junctions at the interface (Fig.~\ref{Fig2}a). This suggests that the observed growth of macroscopic anisotropy reflects a change in shape of each individual micro-junction, rather than an anisotropic modification of their spatial organisation along the contact plane. Indeed, in the latter case, the characteristic length scales measured would have been larger, reflecting the size of possible clusters of micro-junctions. 

To check this hypothesis, we measure $\left|\delta_i-\bar{\delta}\right|$, the displacement of micro-junction $i$ with respect to the average motion of all micro-junctions, $\bar{\delta}$, at the onset of sliding (when $Q=Q_s$). Note that $\bar{\delta}$ varies from about 50 to 150$\mu$m when the normal force is increased from about 1 to 6.4N. $\left|\delta_i-\bar{\delta}\right|$ quantifies how much the initial pattern of micro-junctions along the interface has been modified upon shearing. Figure~\ref{Fig2}b represents $a_i$ as a function of $\left|\delta_i-\bar{\delta}\right|$, for all tracked micro-junctions in all experiments. The large majority of the points are above the equality line, meaning that the micro-junctions move with respect to their initial neighbourhood by less than their own size. Such an observation indicates that the micro-junction pattern forming the real contact is virtually unaffected by shear. Hence, we conclude that the evolution of the correlation length of the real contact does not originate from the relative displacement of micro-junctions. This is in contrast with the anisotropy observed in simulations of frictionless rough viscoelastic contacts~\cite{putignano_viscoelastic_2018}.

To demonstrate that the anisotropic changes in macroscopic correlation lengths originate from an anisotropic change in the shape of the individual micro-junctions, we extract the time evolution of their \rev{equivalent} ellipse (\rev{the ellipse having the same central second-moments as the micro-junction}, see Fig.~\ref{Fig2}c \rev{and \cite{Sahli_SMPRL_2019}}). We define their characteristic dimensions $\ell_{\parallel i}$ and $\ell_{\perp i}$ as the lengths of the cords passing through the ellipse center \rev{along the directions parallel and orthogonal to shear}, respectively \rev{(note that, due to the angle $\theta_i$ between the major axis of the ellipse and the shear direction, $\ell_{\parallel i}$ and $\ell_{\perp i}$ are different from the major and minor axis length of the ellipse)}. Figure~\ref{Fig2}d shows, for all experiments, the average relative variation in $\ell_{\parallel i}$, $(\ell_{\parallel 0i}-\ell_{\parallel si})/\ell_{\parallel 0i}$, as a function of the average relative variation in $\ell_{\perp i}$, $(\ell_{\perp 0i}-\ell_{\perp si})/\ell_{\perp 0i}$, between the initial state (subscript $0$) and that reached when $Q=Q_s$ (subscript $s$). The positive values of both average relative changes show that, under shear, micro-junctions tend to shrink in both directions. The fact that those average points actually lie well above the equality line (by a typical factor of 2-3), indicates that most of the micro-junctions have a larger change \rev{along than orthogonal to the shear direction}. Those results are fully consistent with the observed differential changes in $L_{\parallel}$ and $L_{\perp}$, indicating that the macroscopic growth of anisotropy essentially originates from a gradual shear-induced shape-change of all the individual micro-junctions.

\begin{figure}[ht]
\includegraphics[width=\columnwidth]{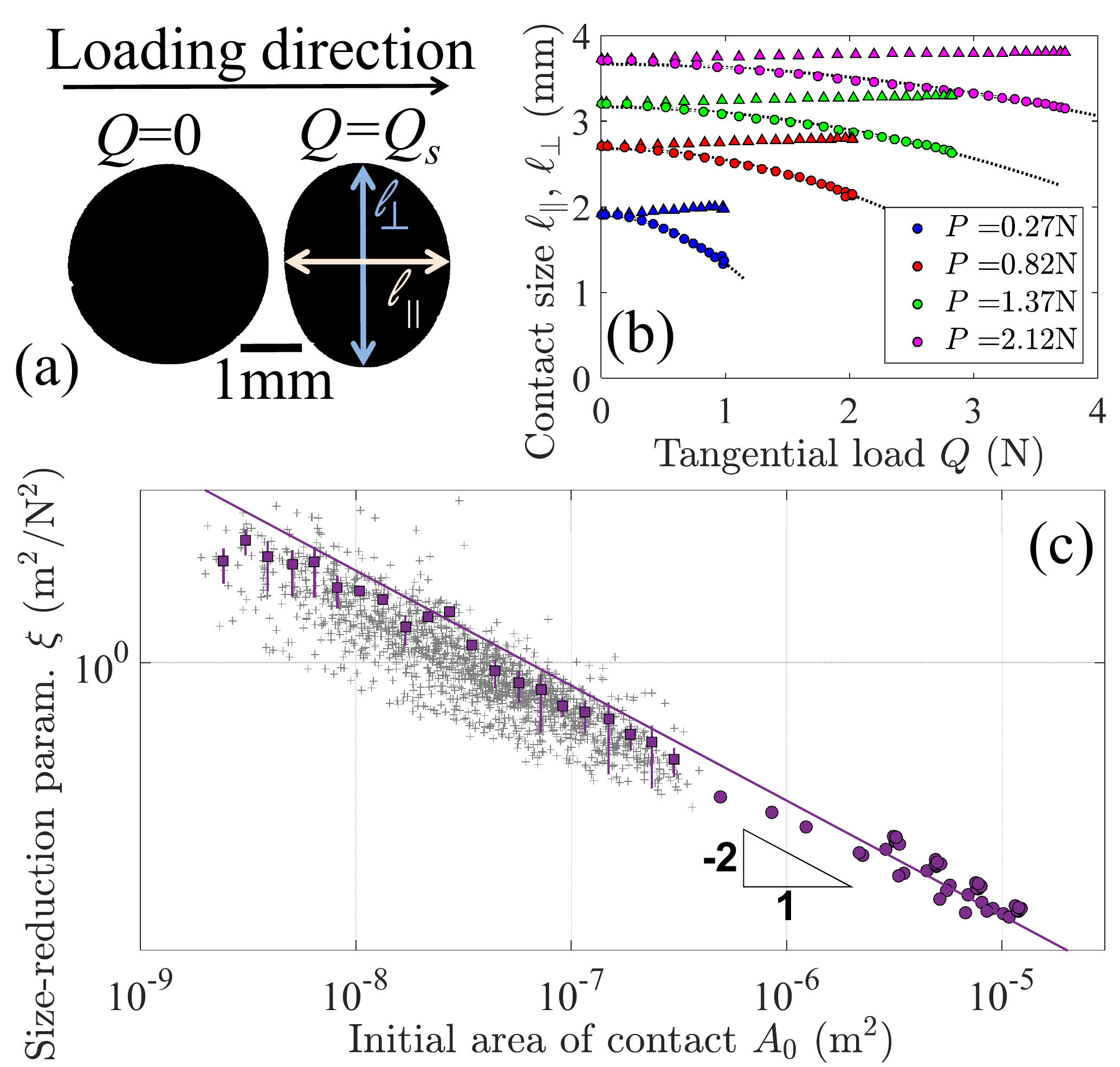}
\caption{(a) Segmented images of a smooth sphere/plane contact ($P$=1.10N, \rev{radius of curvature $R$=9.42mm}), for $Q$=0 and $Q_s$. Right: definition of $\ell_{\parallel}$ and $\ell_{\perp}$. (b) Evolution of $\ell_{\parallel}$ ($\circ$) and $\ell_{\perp}$ ($\triangle$) \textit{vs} $Q$, for smooth sphere/plane contacts, for various normal forces. Dotted lines: fits of the form of Eq.~(\ref{Eq1}). (c) Size-reduction parameter $\xi$ \textit{vs} $A_0$. Disks: smooth sphere/plane contacts. $R$=7.0, 9.42 or 24.0mm. Gray crosses: raw data for micro-junctions within multi-contacts. Squares: average of raw data divided into 21 classes. Error bars: standard deviation within each class. Solid line: guide for eyes, slope -2.\label{Fig3}}
\end{figure}

Understanding the shear-induced anisotropy of rough contacts thus amounts to understanding the shape-changing behaviour of individual micro-junctions. While this shape change is expected to be related to the local tangential force that applies on a micro-junction, the latter force is not a measurable quantity, which impairs direct investigation of the local relationship between aspect ratio and shear force. In order to get some insight about this relationship, we analyze complementary experiments on smooth PDMS-sphere/glass-plane contacts (see~\cite{sahli_evolution_2018} for experimental details). The assumption is that those contacts are good proxies for individual micro-junctions, with the advantage that the tangential force applied on them can be accurately measured. Figure~\ref{Fig3}a shows the evolution of the contact morphology of such a smooth sphere/plane contact under shear. The contact initially has a circular shape, which progressively changes, as the shear force grows, to an ellipse-like shape oriented orthogonal to the shear direction. This is in qualitative agreement with our observations on individual micro-junctions.

Figure~\ref{Fig3}b further shows the evolution of the contact sizes along (resp. orthogonal to) the loading direction, $\ell_{\parallel}$ and $\ell_{\perp}$ respectively, as functions of the tangential force $Q$. For all normal forces, $\ell_{\perp}$ only shows small variations, while the larger variations of $\ell_{\parallel}$ are well captured by an empirical quadratic function of the form (dotted lines)
\begin{equation}
\ell_{\parallel}(Q)=\ell_{\parallel 0}-\xi Q^2, \label{Eq1}
\end{equation}
with the fitting parameter $\xi$ being \textit{a priori} dependent on all system parameters other than $Q$. One of the parameters that we could both change and monitor systematically is the initial area of the contact, $A_0$. Figure~\ref{Fig3}c (disks) shows the evolution of $\xi$ as a function of $A_0$ for all experiments on smooth sphere/plane contacts. We find that $\xi(A_0)$ is well captured by a power law $\xi \sim {A_0}^{\delta}$, with $\delta\simeq$-2.

We now come back to rough contacts and assume that each micro-junction behaves according to Eq.~(\ref{Eq1}): $\ell_{\parallel i}(q_i)=\ell_{\parallel 0i}-\xi_i q_i^2$, with $q_i$ the tangential force on micro-junction $i$. Following \cite{sahli_evolution_2018}, we further assume that, at the onset of macroscopic sliding (index $s$), \textit{i.e.} when $A_{i}=A_{si}$, then $q_i=q_{si}=\sigma A_{si}$, with $\sigma$=0.23MPa the frictional shear strength of the rough PDMS/glass interface. We can thus estimate $\xi_i$ as: $\xi_i=\frac{\ell_{\parallel 0i}-\ell_{\parallel si}}{\sigma^2 {A_{si}}^2}$. In practice, $\ell_{\parallel 0i}$ and $\ell_{\parallel si}$ (resp. $A_{0i}$ and $A_{si}$) are estimated as the initial and final values of a sigmoid fitted onto the time evolution of $\ell_i$ (resp. $A_i$). The resulting values of $\xi_i$ are plotted as a function of $A_{0i}$, as squares in Fig.~\ref{Fig3}c. Strikingly, the micro-junction data align with the smooth sphere/plane contact data (disks). This suggests a common behaviour, for the growth rate of anisotropy under shear, over about four orders of magnitude in $A_0$, from micrometer-scaled micro-junctions within multicontacts to millimeter-scaled sphere/plane contacts. We thus expect the same physical origin for the behaviours observed at both ends of the scale range. In this context, understanding the shear-induced anisotropy of smooth sphere/plane contacts appears as the first step to unravel the anisotropy of sheared rough contacts.

From an empirical standpoint, let us compare the law identified here for the growth of anisotropy of smooth sphere/plane contacts (Eq.~(\ref{Eq1})) and that found for the concurrent reduction of the contact area, $A$. In~\cite{sahli_evolution_2018}, we found that $A=A_0-\alpha Q^2$, with $\alpha \sim {A_0}^{\gamma}$ and $\gamma \simeq -3/2$. Both laws are reminiscent of the ones found above for $\ell_{\parallel}$ and $\xi$, respectively. In order to relate the two sets of observations, we first note that for sphere/plane contacts, $\ell_{\perp}$ remains roughly unchanged under shear, thus suggesting that $\ell_{\perp} \simeq \ell_{\perp 0}$ at all times. Further assuming that the contact takes an elliptic shape, the contact area can thus be written as $A=\frac{\pi}{4} \ell_{\parallel} \ell_{\perp} \simeq \frac{\pi}{4} \ell_{\parallel} \ell_{\perp 0}$. Replacing $A$ in the area reduction law, dividing by $\frac{\pi}{4} {\ell_{\perp 0}}$ and remembering that $\ell_{\parallel 0}=\ell_{\perp 0}=2\sqrt{A_0 / \pi}$, we obtain $\ell_{\parallel}=\ell_{\parallel 0}-\frac{2\alpha}{\sqrt{\pi A_0}}Q^2$. The latter expression both shows that (i) the area reduction law and the anisotropy growth law (Eq.~(\ref{Eq1})) are fully compatible and that (ii) $\xi=\frac{2\alpha}{\sqrt{\pi A_0}}$, thus explaining why the exponents $\delta \simeq -2$ and $\gamma \simeq -3/2$ are found to be related by $\delta=\gamma-1/2$. In this respect, the present results on shear-induced anisotropy are in good agreement with previous results on contact area reduction under shear in the same systems~\cite{sahli_evolution_2018}. They further suggest that the evolution of the area of real contact, $A$, and thus the value of the static friction force (which is proportional to $A$), are actually collateral effects of the shear-induced growth of anisotropy described here.

\rev{A physics-based model of our experiments on smooth sphere/plane contacts can be found in the companion article~\cite{papangelo_shear_2019}, which introduces the first fracture-based contact mechanics model accounting for shear-induced anisotropy in adhesive contacts. Once calibrated on one among the present experiments, that model allows to quantitatively capture the evolution of the contact shape in all other experiments without anymore adjustable parameter. In the model of~\cite{papangelo_shear_2019}, a vanishing work of adhesion between the contacting surfaces corresponds to an absence of evolution of the contact shape under shear, which suggests that adhesion is likely responsible for the experimentally observed shear-induced anisotropy.}

\rev{One interesting aspect of the model of~\cite{papangelo_shear_2019} is that} it can not only be used on initially circular contacts, like those relevant for the present sphere/plane experiments, but also on initially elliptic contacts with a major axis either parallel or perpendicular to the shear loading direction. \rev{The model predicts that} when shear is applied along the major axis, the ellipse's eccentricity tends to decrease, while it increases when shear is orthogonal to the major axis~\cite{papangelo_shear_2019}. Those observations are consistent with the fact that $\ell_{\parallel}$ decreases more than $\ell_{\perp}$ varies under shear. Are those results relevant to rough interfaces, in which micro-junctions have a broad distribution of shapes? To test this, we come back to the \rev{equivalent} ellipse for each micro-junction, already used in Fig.~\ref{Fig2}. We consider all 514 tracked micro-junctions with an initial area larger than 2.10$^{-9}$m$^2$ in the contact under $P$=6.40N. We then calculate Spearman rank correlation coefficient~\cite{spearman_proof_1904} between (i) the set of absolute values of the initial angles, $\left|\theta_i\right|$ (Fig.~\ref{Fig3}c), between the shear direction and the major axis of all micro-junctions and (ii) the corresponding set of relative changes in eccentricities between $Q$=0 and $Q$=$Q_s$. We find a correlation coefficient of -0.36 with a $p$-value of less than 10$^{-16}$, which indicates a significant anticorrelation between both quantities. In other words, angles $\theta_i$ close to 0$^\circ$ (resp. $\pm$90$^\circ$) statistically correspond to decreasing (resp. increasing) eccentricity under shear, in agreement with the theoretical results.

Overall, our results demonstrate that macroscopic rough elastic contacts can develop significant anisotropy under shear, although the topographies and material properties are isotropic. Such anisotropy develops as soon as  shear is applied, well before macroscopic sliding. It originates from a shape change of each micro-junction within the interface, presumably due to the existence of adhesive stresses at the interface. Note that, as already discussed in ~\cite{sahli_evolution_2018}, viscoelasticity is not a likely candidate mechanism, because it cannot be responsible for the sustained anisotropy observed in steady sliding, when strains in the elastomer are no longer time-dependent. We emphasize that the anisotropy that we describe is essentially reversible, in the sense that separating the two solids and performing again a shear experiment will lead to the exact same behaviour. This is in strong contrast with the persistent contact anisotropy induced either by wear (\textit{e.g.} asymmetric scars left in the contact zone ~\cite{tiwari_rubber_2016}) or by shear-driven structural changes in the materials (\textit{e.g.} in fault rocks~\cite{collettini_fault_2009} and in metals~\cite{deng_microstructural_2015}), which would act as anisotropy sources in a subsequent sheared contact. Our results pave the way for a possible control of the many physical properties affected by contact anisotropy (see introduction), through application of controlled shear forces on the interface.


\section{Supplemental Material}

\subsection{Power Spectrum Density (PSD) of the rough surfaces used}
\begin{figure}[ht]
\includegraphics[width=\columnwidth]{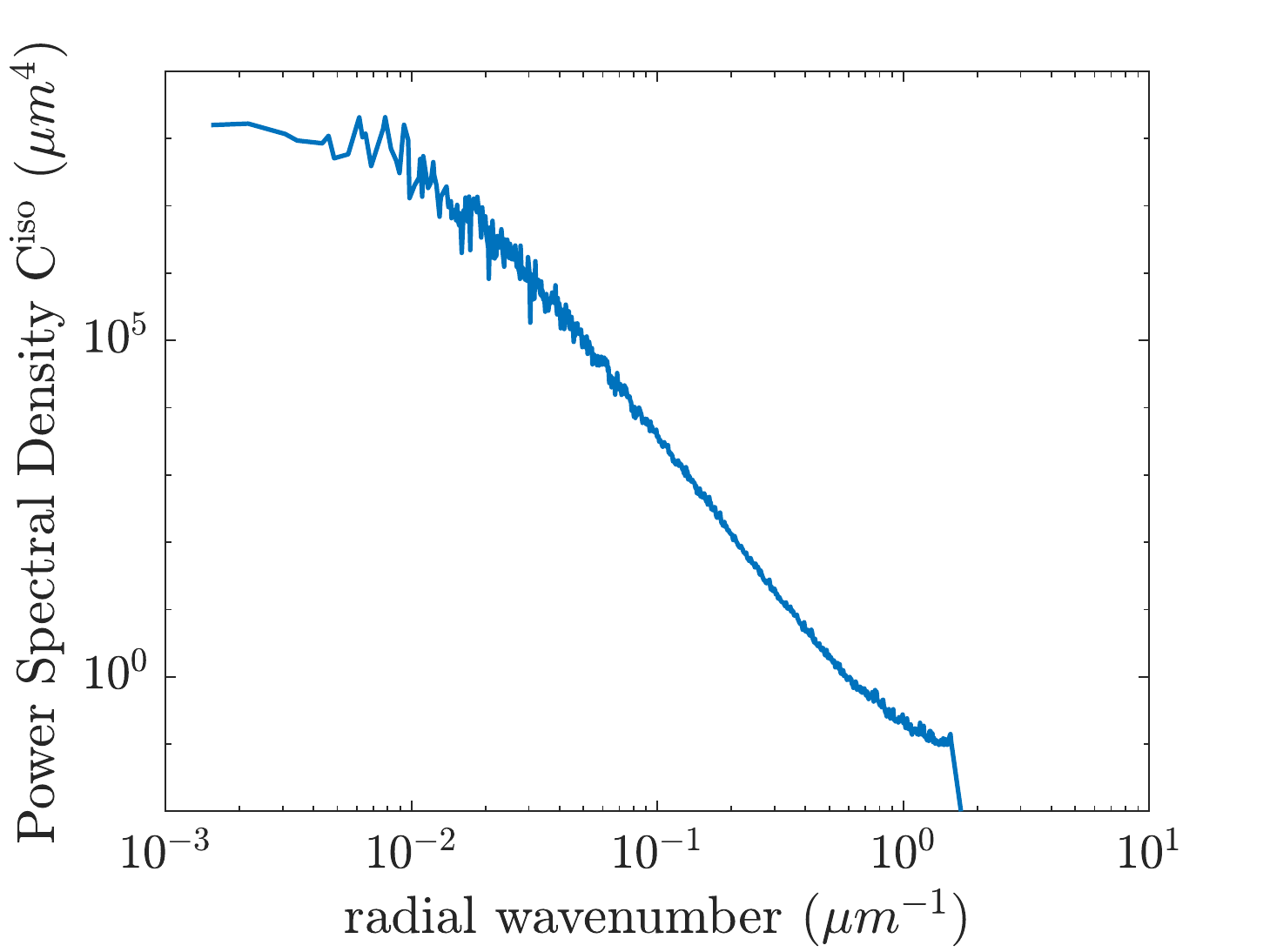}
\caption{Typical PSD ($C^{iso}$-type) of the topography of the rough surfaces used, obtained using the Surface Topography Analyzer (http://contact.engineering/).\label{FigS1}}
\end{figure}

We measured the topography of the rough stainless steel surfaces used to mold the rough PDMS surfaces. The measurement was performed using a stylus profilometer (Surfascan Somicronic) with a 2$\mu$m radius diamond tip, over a 2mm $\times$ 2mm part of the mold. The topography measurement contains 2048$\times$2048 points with a spatial discretization of 2$\mu$m in both the $X$ and $Y$ directions. The PSD was obtained using the online Surface topography Analyzer (http://contact.engineering/). It corresponds to the $C^{iso}$-type of PSD, as defined in \cite{Jacobs_quantitative_2017}. The PSD (Fig.~\ref{FigS1}) features a power-law-like part, of slope about -4.7, limited by two cutoff wavenumbers. The small wavenumber cutoff, $q_0$, corresponds to about 10$^{-2}$ $\mu$m$^{-1}$. Below $q_0$, the PSD is essentially flat. The large wavenumber cutoff is close to the maximum wavenumber accessible ($q_{max}\simeq$2 $\mu$m$^{-1}$) given the spatial resolution of the measurement ($\Delta$x=2$\mu$m). 2$\mu$m is also the radius of curvature of the tip used to probe the topography, so that the large wavenumber cutoff most probably corresponds to the geometrical filtering by the finite-sized probe \cite{lechenault_effects_2010, ponthus_statistics_2019}.

\subsection{Image analysis}
\subsubsection{Threshold determination}
In the raw images of our multicontact interfaces, true contact regions have low grey levels while out-of-contact regions have higher grey levels. Pixels are thus classified based on a threshold whose value is determined automatically, as fully explained in the Supporting Information of \cite{sahli_evolution_2018}, and summarized here.

\begin{figure}[ht]
\includegraphics[width=\columnwidth]{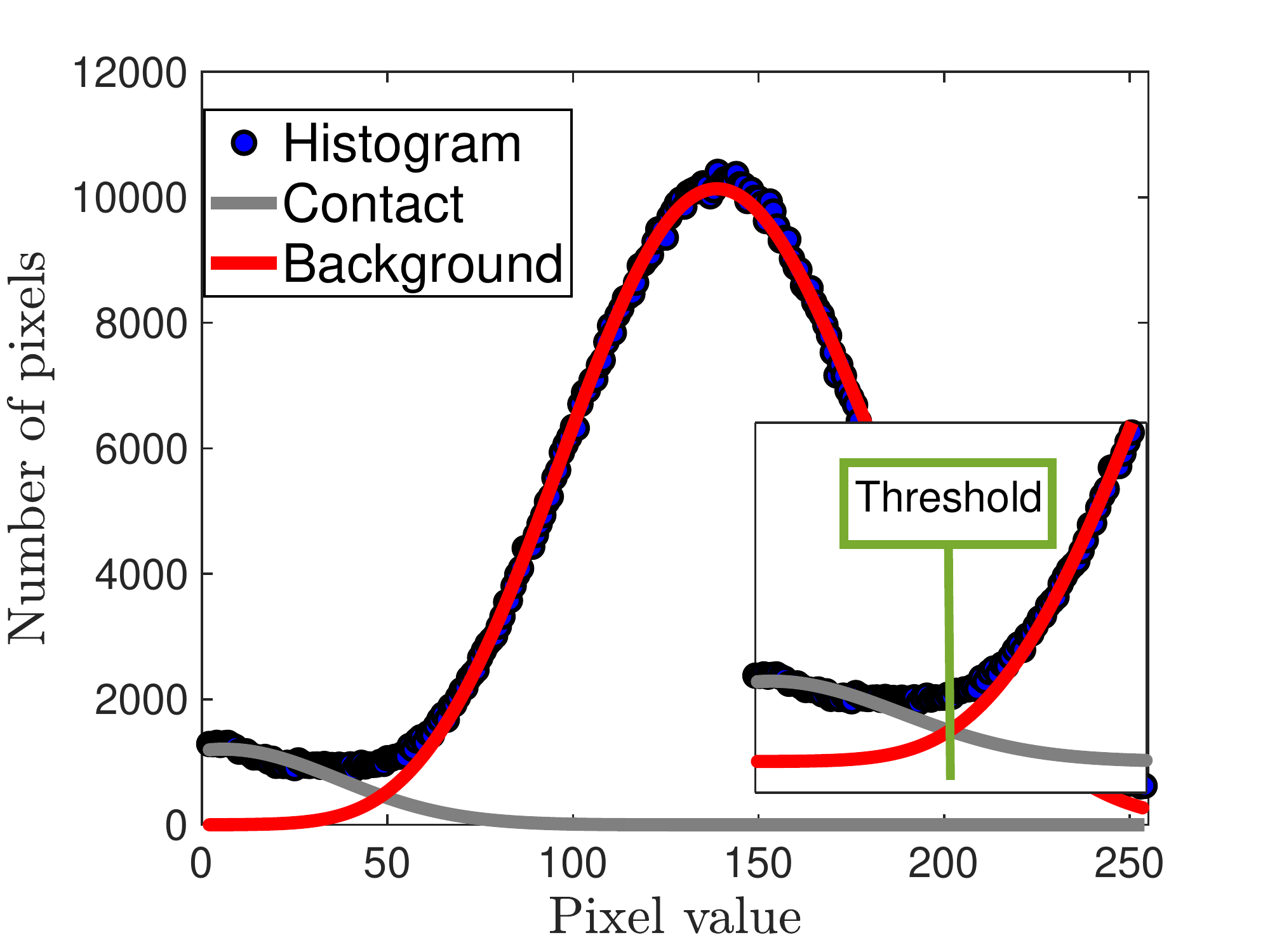}
\caption{Typical image histogram and its fitting with two contributions. Inset: definition of the optimal threshold.\label{FigS2}}
\end{figure}

A typical grey level histogram of a raw image is shown on Fig.~\ref{FigS2}. This is a bimodal histogram, with a week mode located at small grey levels corresponding to in-contact pixels and a large mode corresponding to out-of-contact ones. This histogram can be considered as a weighted sum of two component densities modeling the grey level distribution of respectively in-contact and out-of-contact pixels. Each component is represented by a parametric model, namely a distorted Gaussian inspired by the histogram of images in absence of contact for out-of-contact pixels (red curve) and a Gaussian for in-contact pixels (grey curve). According to Bayes' rule and the maximum a posteriori criterion, the optimal threshold is obtained at the intersection between the two component densities. In practice, it is obtained by (i) identifying the parameters of the mixture model through a least square fitting of the histogram and (ii) analytically solving the intersection equation.


For each normal load, the applied threshold was taken as the mean value of the individual thresholds, determined as above, of all images along the shearing experiment. This average threshold allowed to obtain segmented images like Fig.1b in the main text.

\begin{figure}[h!]
\includegraphics[width=\columnwidth]{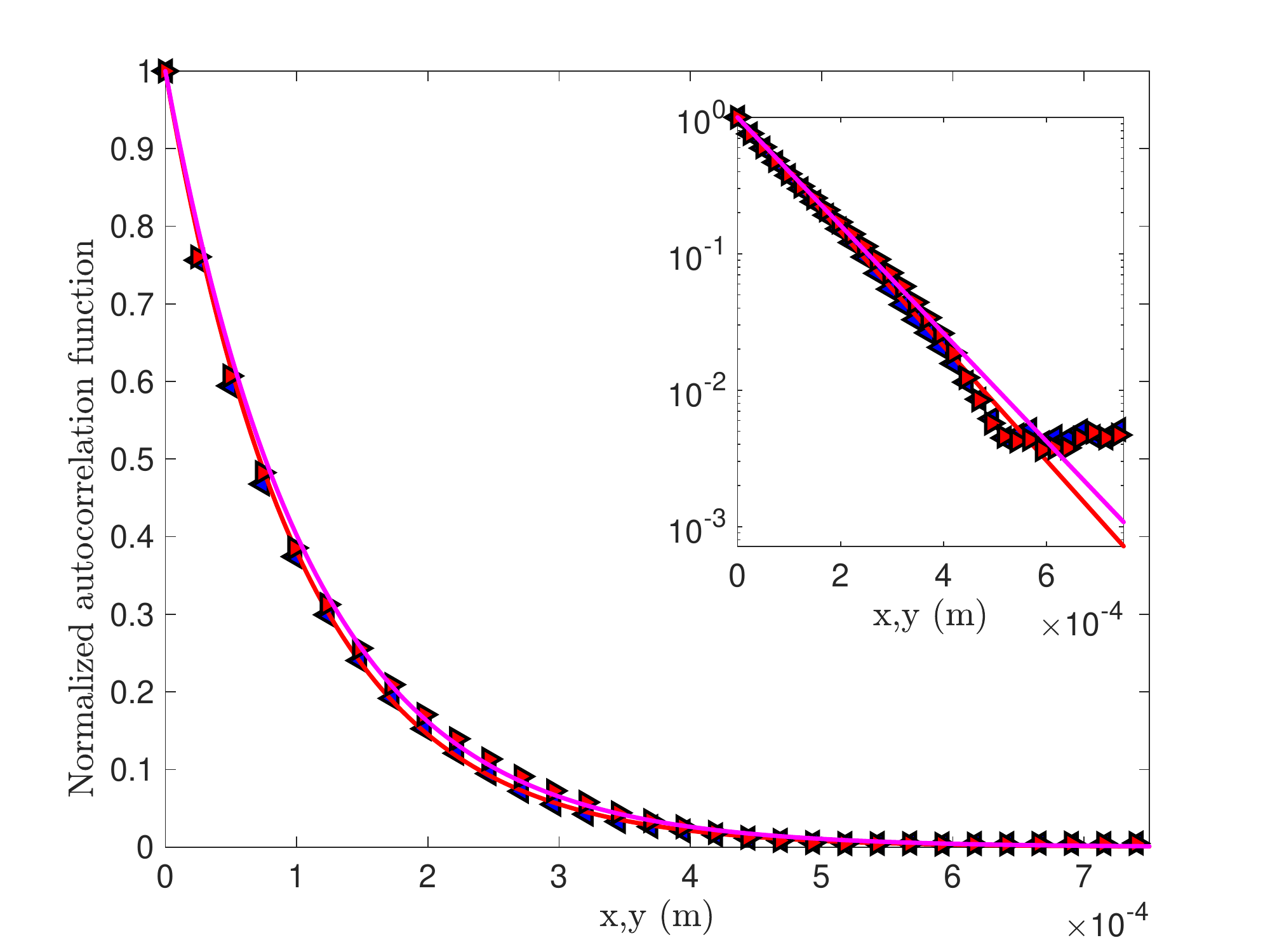}
\caption{Autocorrelation function of the segmented image corresponding to $P$=0.98N and $Q$=1.32N. Blue (red) triangles: cut of the autocorrelation function, through its center, along the direction parallel (perpendicular) to the shear direction. Red (purple) solid line: corresponding cut of the exponential fit of the 2D autocorrelation function (see formula in the caption of Fig.1b) along the direction parallel (perpendicular) to the shear direction, used to estimate $L_{\parallel}$ ($L_{\perp}$). Inset: semilog representation of the same data.\label{FigS3}}
\end{figure}

\subsubsection{Extraction of individual microjunctions' properties}
Individual microjunctions are identified as the connected components of the segmented images using the \textit{regionprops} function in Matlab. The connected components are sets of pixels which are connected by at least one neighbor. We used the 8-connectivity which considers as connected two pixels having at least one common edge or corner.

The term "equivalent ellipse" used in the main text corresponds, for each individual microjunction, to the ellipse having the same central second order moments as the microjunction. In particular, it has the same centroid and area as the microjunction. The eccentricity, angle and axis-lengths used were obtained using the \textit{Eccentricity}, \textit{Orientation}, \textit{MajorAxisLength} and \textit{MinorAxisLength} properties of the \textit{regionprops} function in Matlab.

\subsection{Fit of the autocorrelation function}

Figure~\ref{FigS3} shows the cuts of a typical autocorrelation function of a segmented image along the directions parallel (line $y$=0) and orthogonal to shear (line $x$=0). The linear trend observable in semilog representation (inset) justifies the relevance of the choice of an exponential fit (solid lines) to extract the correlation lengths $L_{\parallel}$ and $L_{\perp}$.\\
\\
\begin{acknowledgments}
This work was supported by LABEX MANUTECH-SISE (ANR-10-LABX-0075) of Universit\'e de Lyon, within the program Investissements d'Avenir (ANR-11-IDEX-0007) operated by the French National Research Agency (ANR). It received funding from the People Program (Marie Curie Actions) of the European Union's Seventh Framework Program (FP7/2007-2013) under Research Executive Agency Grant Agreement PCIG-GA-2011-303871. We are indebted to Institut Carnot Ing\'enierie@Lyon for support and funding. A.P. is thankful to the DFG (German Research Foundation) for funding
the project PA 3303/1-1. M.C. is supported by the Italian Ministry of Education, University and Research (MIUR) under the “Departments of Excellence” grant L.232/2016.
\end{acknowledgments}

\end{document}